\documentclass[pre,twocolumn,showpacs]{revtex4}
\usepackage{epsfig}

\begin{document}

\preprint{MZ-TH/06-07}

\title{Scaling behaviour of non-hyperbolic coupled map lattices}

\author{Stefan Groote}
\affiliation{Teoreetilise F\"u\"usika Instituut, Tartu \"Ulikool,
T\"ahe 4, 51010 Tartu, Estonia\\
and Institut f\"ur Physik der Universit\"at Mainz,
Staudingerweg 7, 55099 Mainz, Germany}
\email{groote@thep.physik.uni-mainz.de}
\author{Christian Beck}
\affiliation{School of Mathematical Sciences, Queen Mary,
University of London, Mile End Road, London E1 4NS, UK}
\email{c.beck@qmul.ac.uk}

\vspace{2cm}

\begin{abstract}
Coupled map lattices of non-hyperbolic local maps arise naturally
in many physical situations described by discretised reaction
diffusion equations or discretised scalar field theories. As a
prototype for these types of lattice dynamical systems we study
diffusively coupled Tchebyscheff maps of $N$-th order which
exhibit strongest possible chaotic behaviour for small coupling
constants $a$. We prove that the expectations of arbitrary
observables scale with $\sqrt{a}$ in the low-coupling limit,
contrasting the hyperbolic case which is known to scale with $a$.
Moreover we prove that there are log-periodic oscillations
of period $\log N^2$ modulating the $\sqrt{a}$-dependence of a
given expectation value. We develop a general 1st order perturbation
theory to analytically calculate the invariant 1-point density,
show that the density exhibits log-periodic oscillations in phase space,
and obtain excellent agreement with numerical results.
\end{abstract}


\maketitle

\section{Introduction}
Coupled map lattices (CMLs) as introduced by Kaneko and
Kapral~\cite{kaneko,kapral} are a paradigm of higher-dimensional
dynamical systems exhibiting spatio-temporal chaotic behaviour.
There is a variety of applications for CMLs to model
hydrodynamical flows, turbulence, chemical reactions, biological
systems, and quantum field theories (see e.g.\ reviews
in~\cite{kanekobook,beckbook}). The analysis of chaotic CMLs is
often restricted to numerical investigations and only a few
analytical results are known. A notable exception is the case of
hyperbolic maps (maps for which the absolute value of the slope
of the local maps is always larger than $1$) for small coupling
$a$. In this case a variety of analytical results
exists~\cite{baladi,jarvenpaa,keller,bunimovich} that guarantee
the existence of a smooth invariant density and ergodic behaviour.
The situation is much more complicated for non-hyperbolic maps
which correspond to the generic case of physical interest. Here
much less is known analytically, though some promising steps have
been made~\cite{chate,ding,ruffo,mackey,beck,gallas}.

In this paper, we present for the first time analytical results
corresponding to a non-hyperbolic situation and calculate
invariant densities explicitely for the case of locally fully
developed chaos that is diffusively coupled. We study local maps
given by $N$-th order Tchebyscheff polynomials. In the uncoupled
case these maps are conjugated to a Bernoulli shift of $N$
symbols. In the coupled case, this conjugacy is destroyed and the
convential treatments for hyperbolic maps does not apply, since
the Tchebyscheff maps have $N-1$ critical points where the slope
vanishes, thus corresponding to a non-hyperbolic situation. From
the physical point of view, the non-hyperbolic case is the most
interesting one. For example, it has been shown that these types
of non-hyperbolic CMLs naturally arise from stochastically
quantised scalar field theories in the anti-integrable
limit~\cite{beckbook}. They can serve as useful models for vacuum
fluctuations and dark energy in the universe~\cite{dark}. Other
applications include chemical kinetics as described by
discretised reaction diffusion dynamics~\cite{kanekobook}.

The case of two coupled Tchebyscheff maps was previously studied
in~\cite{dettmann} using periodic orbit theory. Here we consider
infinitely many diffusively coupled Tchebyscheff maps on
one-dimensional lattices with periodic boundary conditions, and
apply Perron-Frobenius and convolution operator techniques. Our
analytical techniques will yield explicit perturbative expressions
for the invariant 1-point density for small couplings $a$. We will
prove that the density exhibits log-periodic oscillations of
period $\log N^2$ near the edges of the interval. Our explicit
result for the invariant density will allow us to calculate
expectations of arbitrary local observables. We will prove that
expectations of typical observables scale with $\sqrt{a}$ (rather
than with $a$ as for hyperbolic coupled maps). We also show that
there are log-periodic modulations of expectation values when the
parameter $a$ is changed. Our results seem to be typical for
local maps with one or several quadratic maxima which are in a
fully developed chaotic state. Other types of maps may of course
generate different types of behavior for $a\to 0$ \cite{ruffo}.

This paper is organised as follows: In Sec.~II we introduce the
relevant class of coupled map lattices and briefly explain their
physical relevance. In Sec.~III we give some numerical results for
the scaling behaviour of these non-hyperbolic systems. In Sec.~IV we
present our analytical results for the invariant density which we
use to rigorously prove the scaling behaviour.

\section{The class of systems}
Consider a scalar field $\varphi(\vec x,t)$ described by an equation
of the form
\begin{equation}\label{sto2}
\frac{\partial}{\partial t}\varphi=D\cdot\Delta\varphi+
V'(\varphi).
\end{equation}
Here $D$ is a diffusion constant and $V$ is some suitable
potential. These types of equations occur in many different areas
of physics. They describe reaction diffusion systems,
Ginzburg-Landau type of models, nonlinear Schr\"odinger
equations, stochastically quantised field theories, etc. In the
one-dimensional case the Laplacian $\Delta$ is just given by
$\partial^2/\partial x^2$. In many cases there is a fundamental
length and time scale below which the above continuum theory is
not valid anymore. For example, for stochastically quantised
field theories this is the Planck length~\cite{beckbook}. Writing
$t=n\tau$, $x=i\delta$ where $n$ and $i$ are integers, and
$\varphi(x,t)=p_{\rm max}\Phi_n^i$ where $\Phi_n^i$ is a
dimensionless field and $p_{\rm max}$ is some constant with the
same dimension as the field, the discretised Eq.~(\ref{sto2}) can
be written in the form
\begin{equation}\label{dyn}
\Phi_{n+1}^i=(1-a)T(\Phi_n^i)+\frac a2(\Phi_n^{i+1}+\Phi_n^{i-1})
\end{equation}
where the local map $T$ is given by
\begin{equation}\label{map}
T(\Phi)=\Phi+\frac{\tau}{(1-a)p_{\rm max}}V'(p_{\rm max} \Phi),
\end{equation}
and the dimensionless coupling $a$ is given by
$a=2D\tau/\delta^2$. For various applications of these types of
coupled map lattices, see~\cite{kanekobook,beckbook}. The
important point is that generically these types of models can
lead to non-hyperbolic local maps $T$. For example, a
$\varphi^4$-theory described by a double well potential
$V(\varphi)$ with a sufficiently strong quartic term leads to
cubic maps with two inflection points where the slope $T'(\Phi)$
vanishes. It is thus important to understand the generic
behaviour of non-hyperbolic coupled map lattices.

We are particularly interested in cases where the local map exhibits
strongest possible chaotic behaviour. The negative 3rd order
Tchebyscheff map
\begin{equation}
\Phi_{n+1}=T_{-3}(\Phi_n)=-4\Phi_n^3+3\Phi_n
\end{equation}
on the interval $\Phi\in [-1,1]$ is such an example. It is
conjugated to a Bernoulli shift of 3 symbols and can be obtained in our
context from the potential
\begin{equation}\label{16}
V(\varphi)=\frac{1-a}\tau\left(\varphi^2-\frac{\varphi^4}{p_{\rm
max}^2} \right)
  +\mbox{\rm const.}
\end{equation}
In a similar way, one can construct potentials that lead to
positive and negative Tchebyscheff maps of arbitrary order
$N$~\cite{beckbook}.

\section{Observed scaling behaviour}
\begin{figure}
\epsfig{file=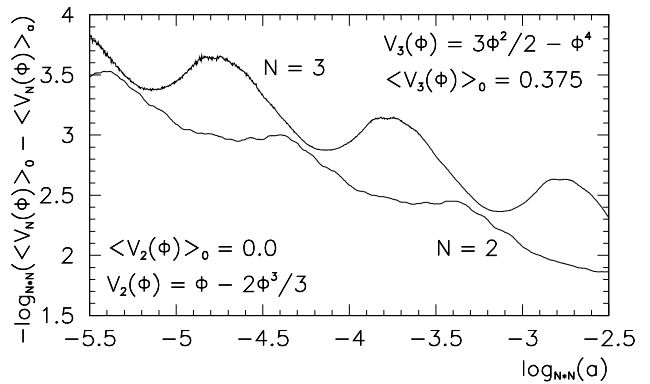, scale=0.5}
\caption{\label{logb1}The function $\sqrt aF^{(N)}(\log a)$ of Eq.~(\ref{a})
with $h(\Phi)=V_N(\Phi)$ as function of $a$ for $N=2,3$ in a
double-logarithmic plot with basis $N^2$.}
\end{figure}

Let us now study CMLs of type~(\ref{dyn}) where $T=T_N$ is a
Tchebyscheff map of order $N$. One observes nontrivial scaling
behaviour for small values of the coupling $a$ that is significantly
different from that of hyperbolic systems. Consider an arbitrary test
function $h(\Phi)$ of the local iterates. Assuming ergodicity,
expectations $\langle h(\Phi)\rangle_a$ for a given parameter $a$ are
numerically calculable as time averages
\begin{equation}
\langle h(\Phi)\rangle_a=\lim_{M\to\infty,J\to\infty}
\frac1{MJ}\sum_{n=1}^M\sum_{i=1}^Jh(\Phi_n^i).
\end{equation}
For $a\to 0$ one numerically observes the scaling behaviour
\begin{equation}\label{a}
\langle h(\Phi)\rangle_a-\langle h(\Phi)\rangle_0
  =\sqrt{a}\cdot F^{(N)}(\log a)
\end{equation}
where $F^{(N)}$ is a periodic function of $\log a$ with period
$\log N^2$. Examples are shown in FIG.~\ref{logb1}.
The choice $h(\Phi)=V_{\pm N}(\Phi):=\mp\int T_N(\Phi)d\Phi$
is important in the physical applications to estimate the
vacuum energy generated by the chaotic field theory under
consideration~\cite{beckbook}. Generally the function $F=F[h]$ is
a functional of the chosen test function $h$.

\begin{figure}
\epsfig{file=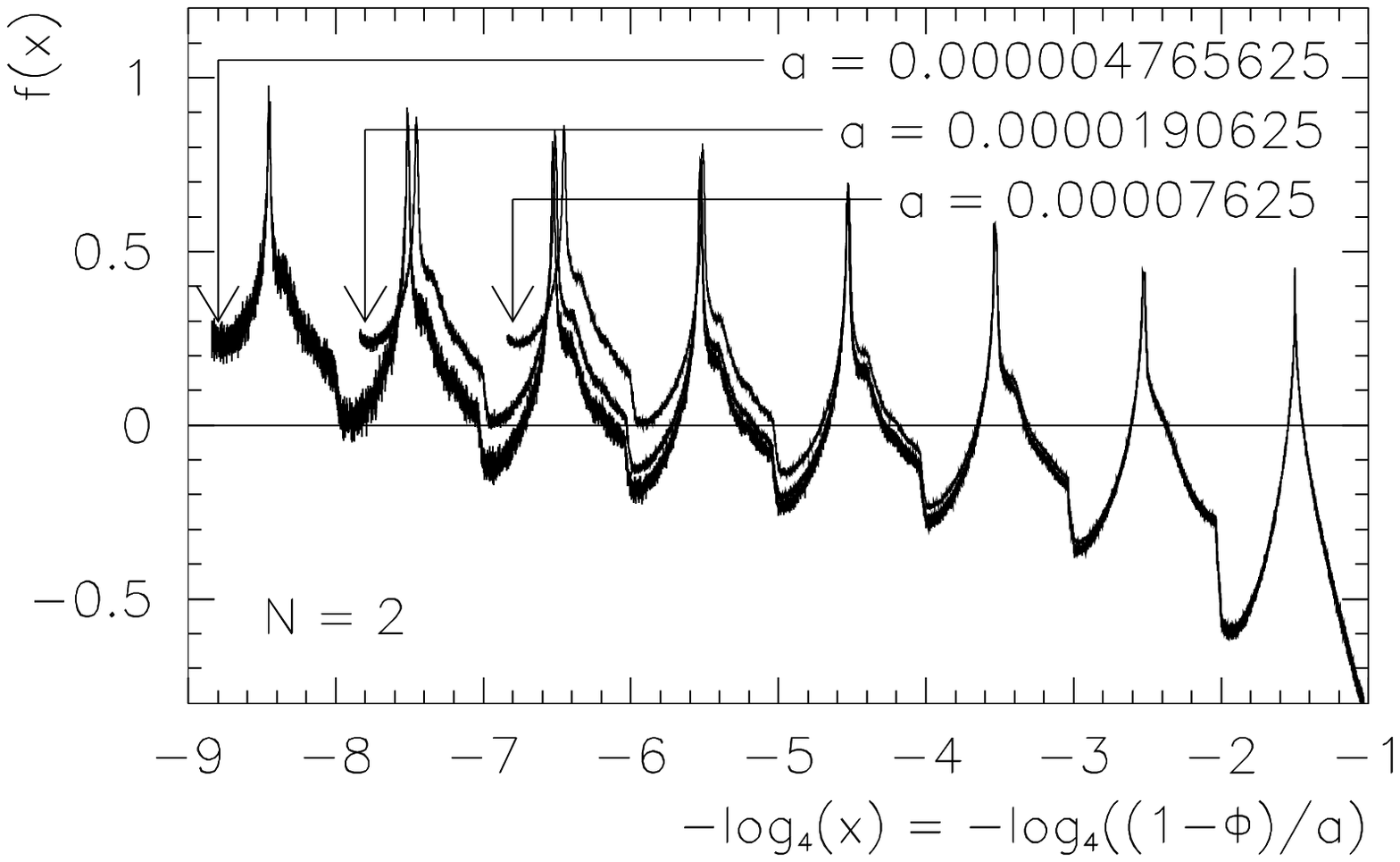, scale=0.5}\\[7pt]
\epsfig{file=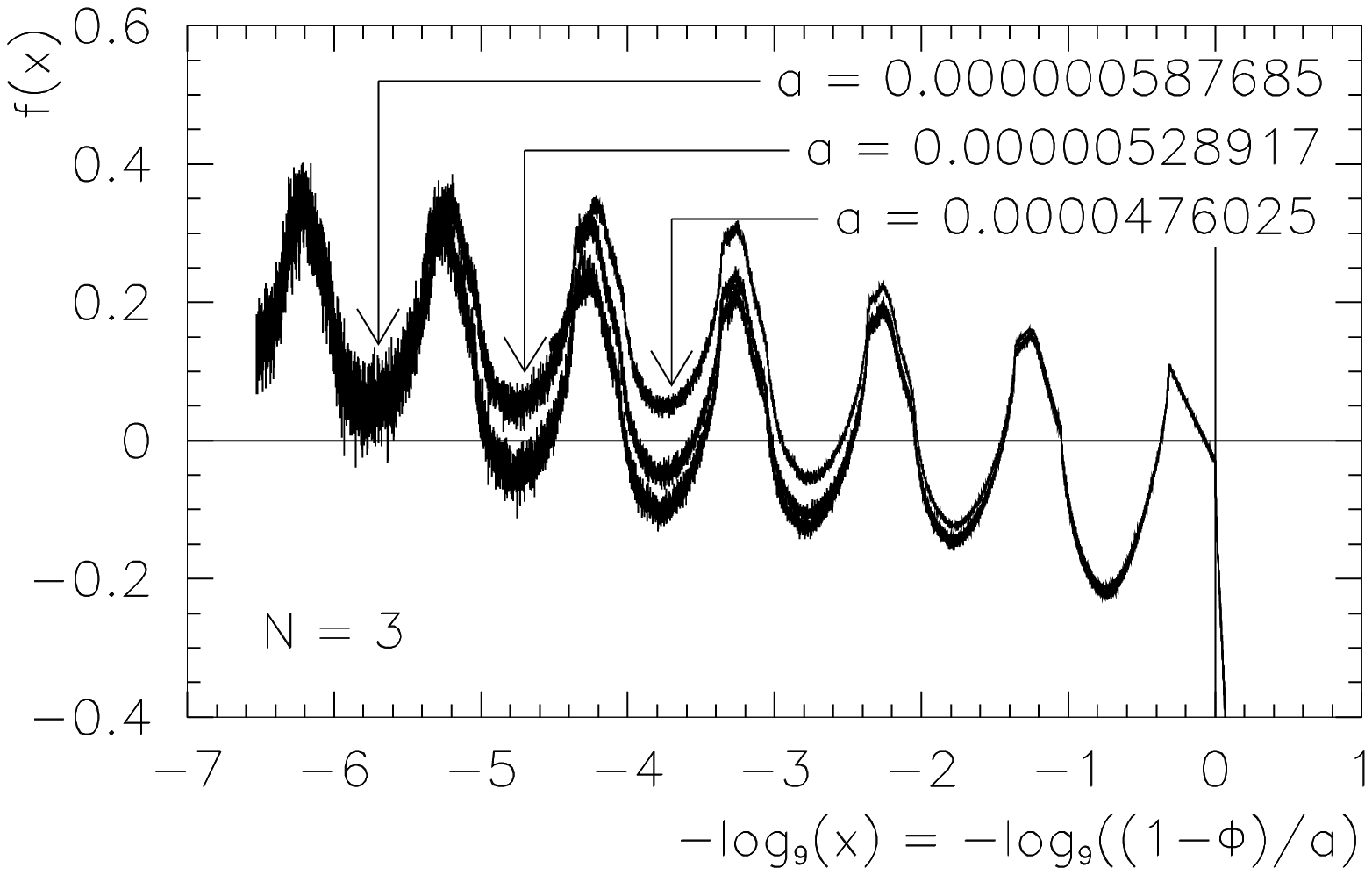, scale=0.5}
\caption{\label{peri}Log-periodic oscillations of the rescaled
invariant density $f(x)$ as observed for $N=2$ (top) and $N=3$
(bottom) and different values for $a$.}
\end{figure}

The above log-periodic scaling is observed for arbitrary test
functions $h$ and hence is a general property of the invariant
1-point density $\rho_a(\Phi)$ of the CML for given small
couplings $a$. In fact one observes that there is not only
scaling in the parameter space $a$ but also in the phase space
$\Phi$. This is shown in FIG.~\ref{peri}.

Near the left edge of the interval $[-1,1]$ one may write $\Phi=ay-1$
and observe the scaling behaviour
\begin{equation}\label{g}
\rho_a(ay-1)=a^{-1/2}g(y)
\end{equation}
where the function $g$ is independent of $a$ for small $a$. At
the right edge, writing $\Phi=1-ax$, one observes
\begin{equation}\label{f}
\rho_a(1-ax)=\rho_0(1-ax)+\frac12a^{-1/2}x^{-1}f(x)
\end{equation}
where $f$ is independent of $a$ for small $a$. Moreover, $f$
exhibits log-periodic oscillations
\begin{equation}\label{ff}
f(N^2x)=f(x)
\end{equation}
over a large region of the phase space (see FIG.~\ref{peri}).
The number of oscillations is approximately $-\log_{N^2}(a)$.

For $N$ odd the density is symmetric and the behaviour at the left
and right edge is the same, whereas for $N$ even there is no
left-right symmetry.

In the following section we will analytically prove the above
numerical observations and provide explicit formulas for the
functions $f$ and $g$ for a given local Tchebyscheff map $T_N$.

\section{Perturbative calculation of the invariant density}
Our method is based on a perturbative treatment of the Perron-Frobenius
operator of a perturbed local map and on convolution techniques. In a
first approximation, the neighbouring lattice sites can be regarded as
producing independent random noise with density
$\rho_0(\Phi)=(1-\Phi^2)^{-1/2}/\pi$. While this picture works well in
the middle of the interval $[-1,1]$, in the vicinity of the edges $\pm 1$
one has to take into account nontrivial nearest neighbour correlations,
due to the non-hyperbolicity of the map. A first-order approximation of
the density can then be further iterated to yield results of better
precision. In each step we integrate over 2-point functions
describing the joint probability of neighboured lattice sites to
obtain the marginal distribution at a single lattice site. A
detailed description of our calculations is out of the scope of a
short letter, they will be published elsewhere~\cite{groote}. Our
final result is that for $N=2$ one obtains in leading order of
$\sqrt{a}$ at the left edge $\rho_a(ay-1)\approx\rho_a^{(0)}(ay-1)$ where
\begin{equation}\label{15}
\rho_a^{(0)}(ay-1)=\frac1{\pi\sqrt{2a}}\int\frac{\rho_0(\phi_+)d\phi_+
  \rho_0(\phi_-)d\phi_-}{\sqrt{y-1+(\phi_++\phi_-)/2}}.
\end{equation}
At the right edge one obtains by iterating our scheme $q$ times
$\rho_a(1-ax)\approx\sum_{p=1}^q\rho_a^{(p)}(1-ax)$ where
\begin{equation}\label{166}
\rho^{(p)}_a(1-ax)=\frac1{4^p\pi\sqrt{2a}}\int
\frac{\rho_0(\phi_+)d\phi_+\rho_0(\phi_-)
  d\phi_-}{\sqrt{x/4^p+r_2^p(\phi_+)+r_2^p(\phi_-)}}.
\end{equation}
Here the function $r_2^p(\phi)$ is defined as follows:
\begin{equation}
r_2^p(\phi)=\frac12\sum_{q=0}^p\frac{T_{2^q}(\phi)-1}{2^{2q}}.
\end{equation}
The limits of the two integrations in Eqs.~(\ref{15}) and~(\ref{166})
are given by the condition that $|\phi_{\pm}|\leq 1$ and that the
argument of the square root should always be positive.

A simple way to numerically evaluate our formulas is to replace the
double integrals by ergodic averages of iterates of the uncoupled
Tchebyscheff map. So for example we may evaluate the density at the
left edge as
\begin{equation}
\rho_a^{(0)}(ay-1)=\frac1{\pi\sqrt{2a}}\frac1{M^2}\!\!\sum_{n_\pm=1}^M
\frac{\theta(y-1+(\Phi_{n_+}+\Phi_{n_-})/2)}{\sqrt{y-1
  +(\Phi_{n_+}+\Phi_{n_-})/2}},
\end{equation}
and similar formulas apply to the right edge.

\begin{figure}
\epsfig{file=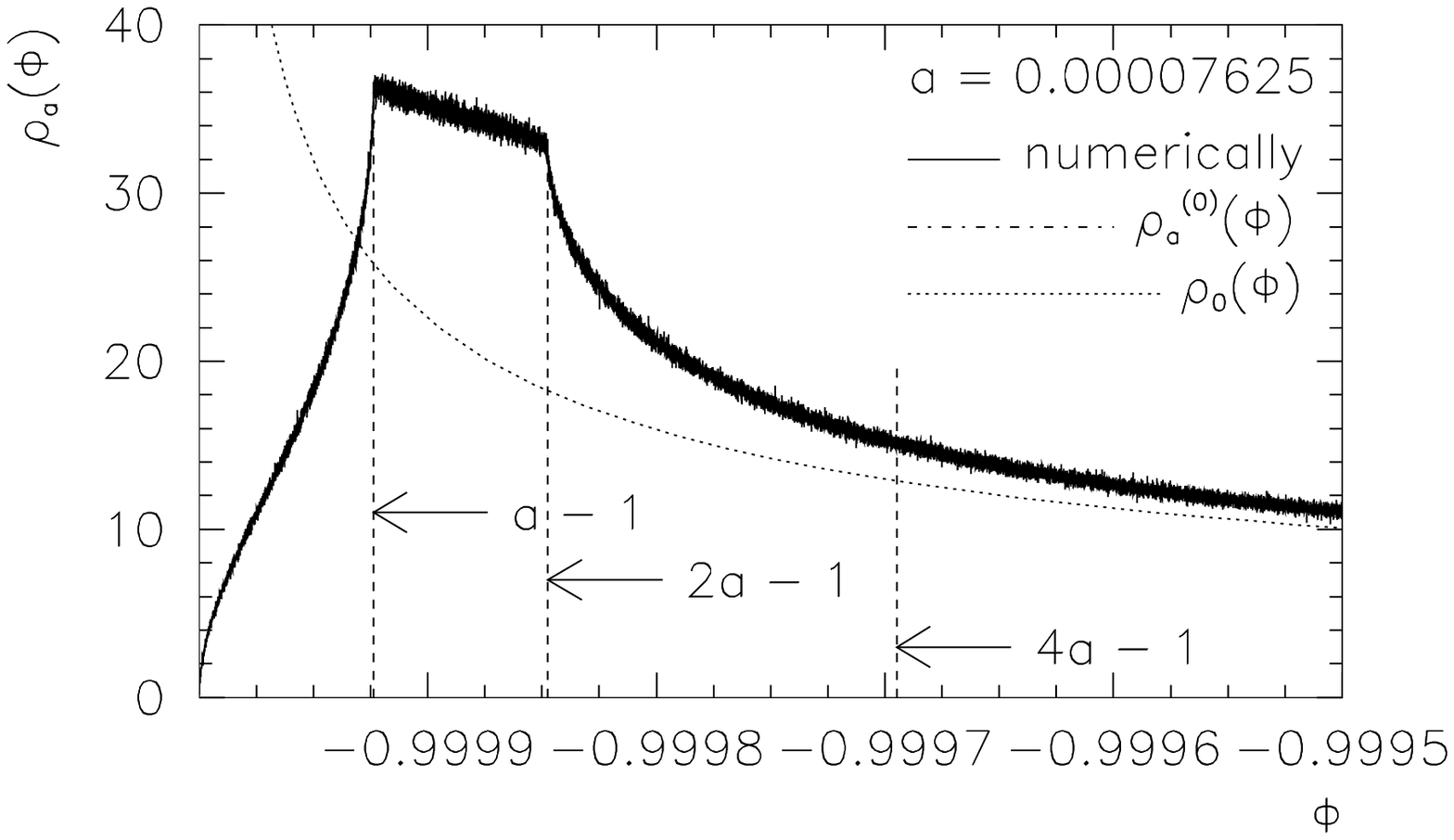, scale=0.5}\\[7pt]\strut\kern-15pt
\epsfig{file=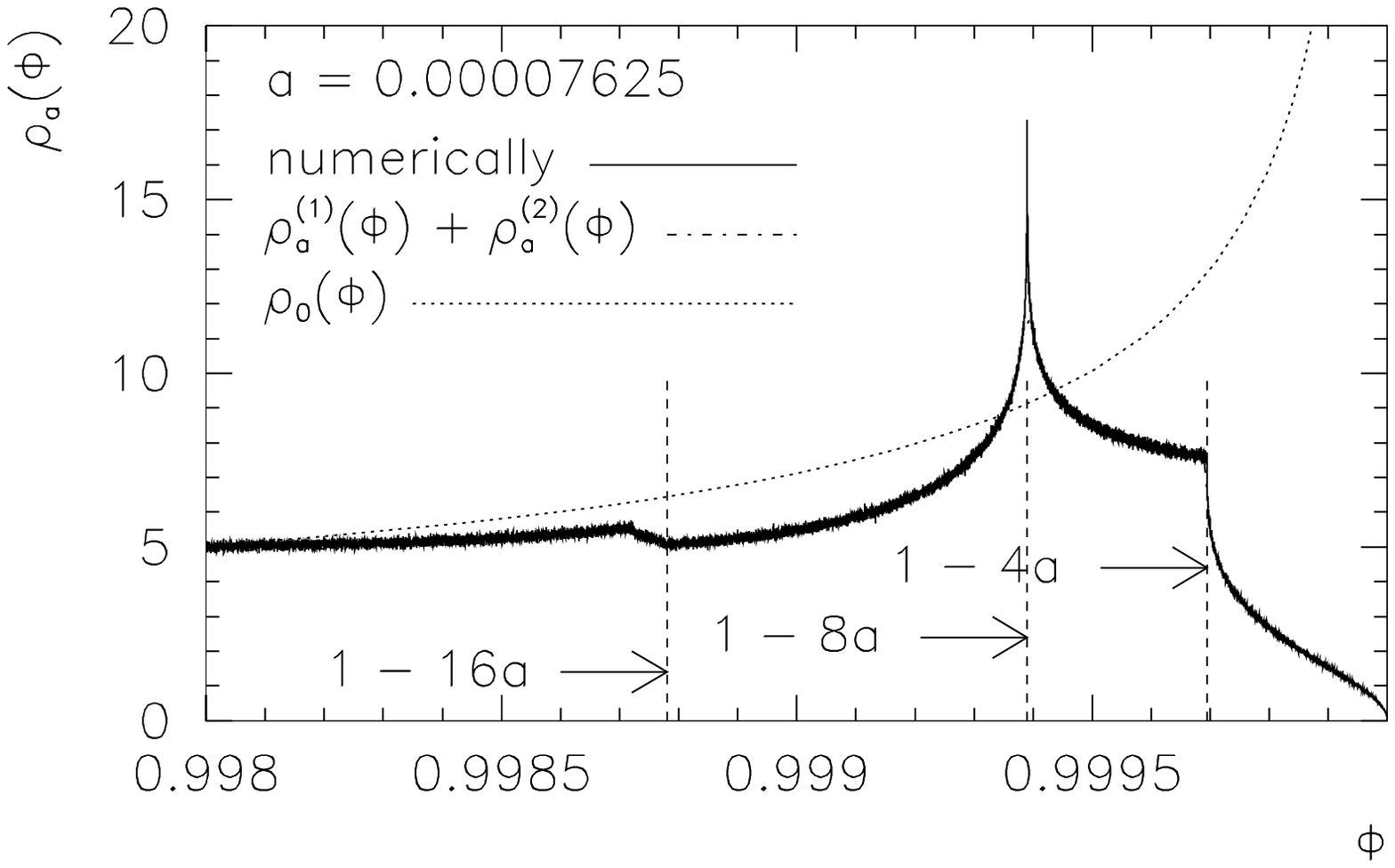, scale=0.5}
\caption{\label{rho23}Scaling behaviour of the invariant 1-point density
  near the left and right edge of the interval [-1,1] for $N=2$ and
  $a=0.00007625$. Shown is the density at the left (upper diagram) and right
  edge (lower diagram) in comparison with the exact results (dashed-dotted
  curve, covered by the numerical result) and the density $\rho_0(\phi)$
  (dotted curve) together with several thresholds, i.e.\ points
  with nonanalytic behavior (dashed lines).}
\end{figure}

Eq.~(\ref{15}) can also be written as
\begin{eqnarray}
\rho_a^{(0)}(ay-1)&=&\frac1{\pi\sqrt{2a}}\int_{1-y}^1
  \frac{\rho_{00}(z)dz}{\sqrt{y-1+z}} \\
\rho_{00}(z)&=&\frac{2}{\pi^2}K(\sqrt{1-z^2})\theta(1-z^2)
\end{eqnarray}
where $K(x)$ is the complete elliptic integral of the first kind.
From the above formula it is obvious that the density is not
differentiable at $y=1$ and $y=2$ for arbitrarily small $a$ (compare
FIG.~\ref{rho23}). By iteration of the Perron-Frobenius operator one
can show that these two non-analytic points generate an entire cascade
of such points at the right edge of the interval.

Note that the sum over the terms in Eq.~(\ref{166}) converges
rapidly with increasing $q$. In practice, a few terms are
sufficient to obtain perfect agreement with the numerical
histograms. FIG.~\ref{rho23} shows how well our analytical
results~(\ref{15}) and~(\ref{166}) agree with the numerics.
The agreement is so good that the analytic curves (given by dashed-dotted
lines in FIG.~\ref{rho23}) are not visible behind the data points.

Along similar lines, we obtain for $N=3$
\begin{equation}
\rho_a^{(p)}(1-ax)=\frac2{9^p3\pi\sqrt{2a}}\int\frac{\rho_0(\phi_+)d\phi_+
  \rho_0(\phi_-)d\phi_-}{\sqrt{x/9^p+r_3^p(\phi_+)+r_3^p(\phi_-)}},
\end{equation}
with
\begin{equation}
r_3^p(\phi)=\frac12\sum_{q=0}^p\frac{T_{3^q}(\phi)-1}{9^q}.
\end{equation}
The density is symmetric, i.e.
\begin{equation}
\rho_a^{(p)}(ax-1)=\rho_a^{(p)}(1-ax).
\end{equation}

For general $N$ it is natural to conjecture that
\begin{equation}
\rho_a^{(p)}(1-ax)\sim\frac{1}{\sqrt{a}}\int\frac{\rho_0(\phi_+)d\phi_+
  \rho_0(\phi_-)d\phi_-}{\sqrt{x/N^{2p}+r_N^p(\phi_+)+r_N^p(\phi_-)}},
\end{equation}
with
\begin{equation}
r_N^p(\phi)=\frac12\sum_{q=0}^p\frac{T_{N^q}(\phi)-1}{N^{2q}}.
\end{equation}

With the above equations we have derived explicit formulas for
the scaling functions $f$ and $g$ in Eqs.~(\ref{g}) and~(\ref{f}).
It is easy to check from this integral representation that $f$
indeed satisfies property~(\ref{ff}). Using Eqs.~(\ref{g}),
(\ref{f}) and~(\ref{ff}) it is also possible to prove the general
scaling relation~(\ref{a}) for arbitrary test functions $h$. Due
to space restrictions we do not describe the details here but refer
to a longer version~\cite{groote}. The function $F^{(N)}$ is
essentially given by a suitable integral of the observable $h$
folded with $f$.

\section{Conclusion and outlook}
We have derived in a perturbative way the 1-point density for
coupled Tchebyscheff maps and rigorously proved the existence of
several interesting scaling phenomena that are numerically
confirmed. Although our formulas were worked out for the explicit
example of Tchebyscheff maps, many of our techniques and results
are expected to hold in a similar way for other non-hyperbolic
systems exhibiting strongly chaotic behaviour. Scaling with
$\sqrt{a}$ and log-periodic oscillations seem to be typical
phenomena for nonhyperbolic dynamical systems for which the local
map has one or several quadratic maxima and is in a fully
developed chaotic state.

\begin{acknowledgments}
This work is supported in part by the Estonian target financed
project No 0182647s04 and by the Estonian Science Foundation under
grant No. 6216. S.G. also acknowledges support from a grant of the
Deutsche Forschungsgemeinschaft (DFG) for staying at Mainz
University as guest scientist for a couple of months. C.B.'s
research is supported by a Springboard Fellowship of the
Engineering and Physical Sciences Research Council (EPSRC).
\end{acknowledgments}

\end{document}